\documentclass{iau} 
\usepackage{graphicx}

\title[The VLT-FLAMES Tarantula Survey] 
{The VLT-FLAMES Tarantula Survey}

\author[Jorick S. Vink et al.,]   
{Jorick S. Vink$^1$, C.J. Evans$^2$, J. Bestenlehner$^{1,3}$, C. McEvoy$^4$, O. Ram\'{i}rez-Agudelo$^{2}$,  H. Sana$^5$, F. Schneider$^{6}$ \and VFTS 
}

\affiliation{$^1$Armagh Observatory, College Hill, BT61 9DG, Armagh, Northern Ireland \\ email: {\tt jsv@arm.ac.uk} 
$^2$ATC, Royal Observatory Edinburgh, Blackford Hill, Edinburgh, EH9 3HJ, UK
$^3$Departament of Physic and Astronomy University of Sheffield, Sheffield, S3 7RH, UK
$^4$ARC, School of Mathematics and Physics, QUB, Belfast BT7 1NN, UK
$^5$Institute of Astrophysics, KU Leuven, Celestijnenlaan 200D, 3001, Leuven, Belgium
$^6$Department of Physics, University of Oxford, Keble Road, Oxford OX1 3RH, UK}

\pubyear{2017}
\volume{329}  
\jname{Title of your IAU Symposium}
\editors{A.C. Editor, B.D. Editor \& C.E. Editor, eds.}
\begin{document}

\maketitle

\begin{abstract}
We present a number of notable results from the VLT-FLAMES Tarantula Survey (VFTS), an ESO Large Program during which we 
obtained multi-epoch medium-resolution optical spectroscopy of a very large sample of over 800 massive stars in the 30 Doradus region of 
the Large Magellanic Cloud (LMC). This unprecedented data-set has enabled us to address some key questions regarding
atmospheres and winds, as well as the evolution of (very) massive stars.
Here we focus on O-type runaways, the width of the main sequence, and the mass-loss rates for (very) massive stars. 
We also provide indications for the presence of a {\it top-heavy} initial mass function (IMF) in 30 Dor.
\end{abstract}

\firstsection 
\section{Introduction}

Massive star evolution is important for many fields of Astrophysics including supernovae (SNe; Levesque, these proceedings). 
Yet, it remains largely unconstrained (Langer 2012; Meynet these proceedings). Progress can be made using
high-quality observations from nearby sources, as well as from large data-sets such as 
VFTS (Evans et al. 2011) discussed here. 
In parallel, VFTS data are analysed using state-of-the-art model atmospheres such as 
CMFGEN (Hiller \& Miller 1998) and FASTWIND (Puls et al. 2005), as well as 
automatic fitting tools (Sab{\'{\i}}n-Sanjuli{\'a}n et al. 2014; Bestenlehner et al. 2014; Ram\'{i}rez-Agudelo et al. 2017).

In addition to this observational progress, our VFTS collaboration strives to make theoretical progress on 
stellar winds and evolution, and we are in the unique position to confront our new models against 
VFTS data. In the following, we highlight a number of recent results that we argue 
make a real difference to our knowledge of massive stars. 

\subsection{Motivation for the Tarantula region}

The Tarantula region (30 Doradus) is the largest active star-forming region in our 
Local Universe for which individual spectra of the massive-star population can be 
obtained. Because it is the largest region, it provides a unique opportunity 
to study the most massive stars, including very massive stars (VMS) with masses up to 200-300 $M_{\odot}$ 
(Crowther et al. 2010; Bestenlehner et al. 2014; Martins 2015; Vink et al. 2015). This  
allows us to properly investigate whether the upper-IMF may be 
top-heavy (Schneider et al. 2017). Answering this question is important as these VMS that 
are thought to dominate the ionizing radiation and wind feedback from massive stars 
(Doran et al. 2013). 

Another reason to study 30 Doradus is that testing massive star evolution
{\it requires} large data-sets. For instance, the issue of the location of 
the terminal-age main sequence (TAMS) can only be addressed when the sample-size is 
sufficiently large to populate both the main-sequence with O-type stars (Sab{\'{\i}}n-Sanjuli{\'a}n et al. 2017; Ram\'{i}rez-Agudelo et al. 2017) 
and B supergiants (McEvoy et al. 2015).   

\section{Results on binarity, rotation rates, and runaways}

\begin{figure}
\begin{center}
\includegraphics[width=\textwidth]{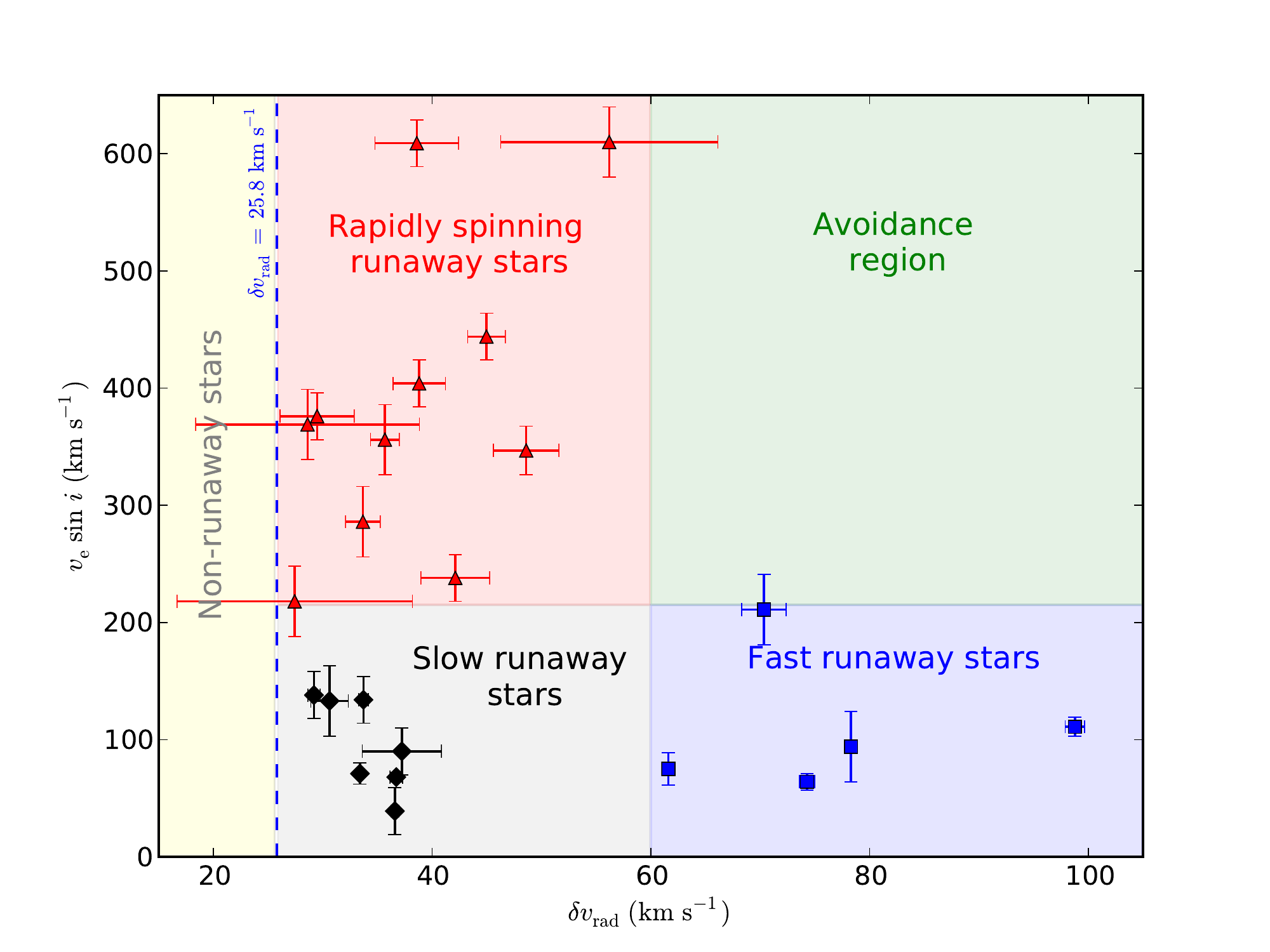}
\caption{Rotational velocities of both slow \& fast runaways (Sana et al. in prep.). 
Whilst there are slow runaways at high $v$ $\sin$ $i$, and fast runaways at relatively 
low $v$ $\sin$ $i$, there appears to be a Region-of-Avoidance for fast runaways with large 
$v$ $\sin$ $i$. These diagnostics might enable us to disentangle the different proposed 
origins for runaways, as discussed in the text.}
\label{f_runaways}
\end{center}
\end{figure}

The aims of VFTS were to determine the stellar parameters, such as  
$T_{\rm eff}$, $\log g$ \& $\log L$ to place our objects on the HR-diagram; the mass and $\dot{M}$ to determine 
the evolution \& fate of massive stars; and the helium (He) and nitrogen (N) 
abundances to test (rotational) mixing processes (Grin et al. 2016; 
Rivero-Gonzalez et al. 2012).
All these parameters require sophisticated atmosphere modeling, but VFTS  
also offered some model-independent parameters including the rotational velocities $v$ $\sin$ $i$ and 
radial velocities (RVs) thanks to the multi-epoch nature of the survey. The latter allowed 
us to obtain information on the $\sim$ 50\% frequency in 30 Dor (Sana et al. 2013) and the opportunity 
to study the dominant mechanism for runaways (Fig.\,\ref{f_runaways}; Sana et al. in prep.).

Figure\,\ref{f_runaways} might allow us to disentangle the dynamical runaway scenario 
(Gies \& Bolton 1986) from the binary-SN kick scenario (Stone 1991), as the first scenario
might produce relatively fast runaways, whilst one would expect the binary SN kick scenario 
to produce rapid rotators. Obviously, definitive conclusions can only be obtained when 
more sophisticated models become available. 

\section{The width of the main-sequence and constraints on core overshooting}

\begin{figure}
\begin{center}
\includegraphics[width=\textwidth]{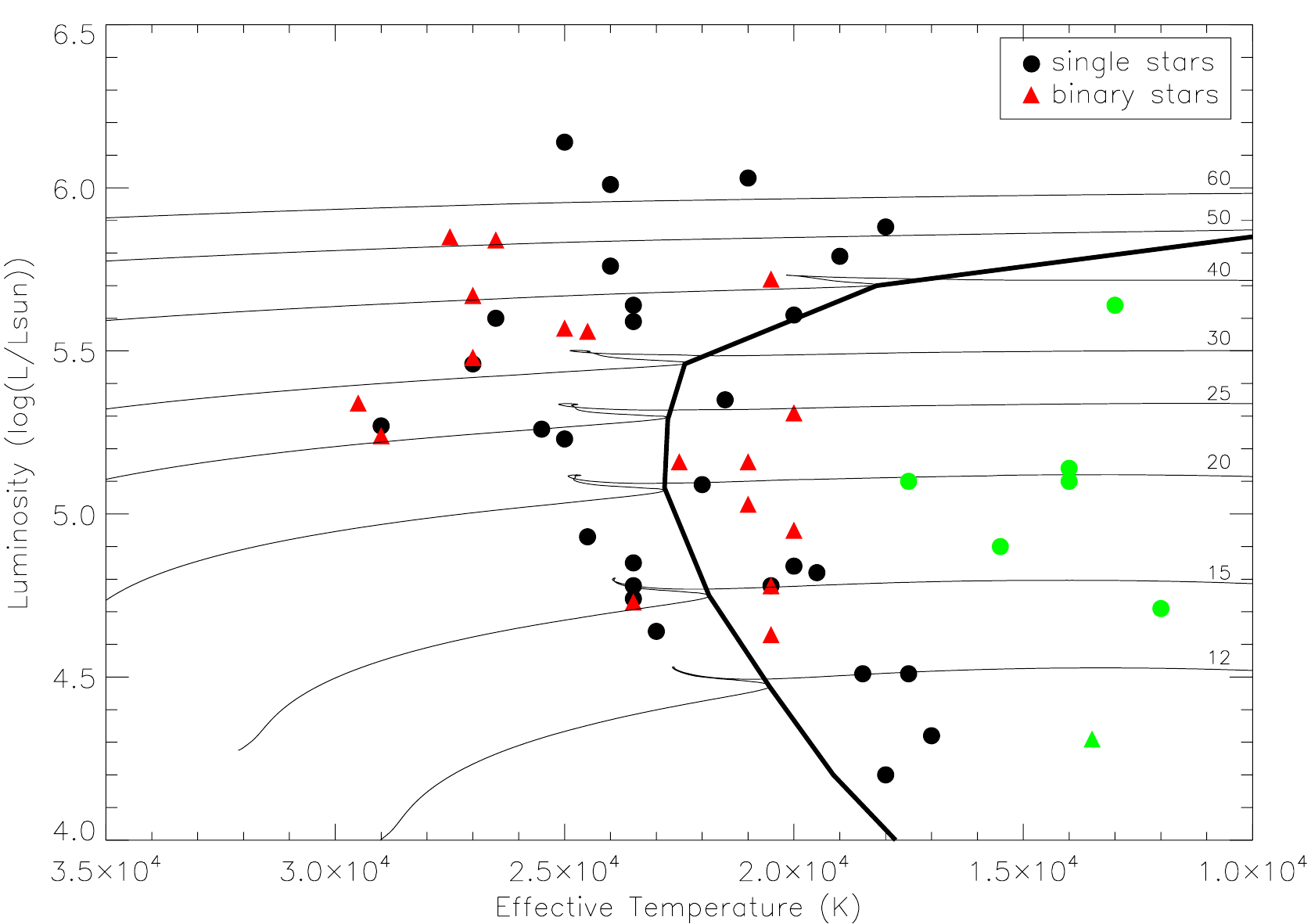}
\caption{A zoomed-in version of the Hertzsprung-Russell diagram for
  single (circle) and binary (triangle) B supergiants from McEvoy et al. (2015). Also shown
  are the LMC evolutionary tracks of Brott et al. (2011) for $v_{\rm rot}$ 
  $\simeq 225$\,km/s, with the initial mass (in units of the solar mass) given
  on the right hand side. The dark line represents the TAMS (terminal
  age main sequence). 
  The stars highlighted in green are far enough
  from the TAMS line that they may be interpreted as core He-burning objects.}
\label{f_TAMS}
\end{center}
\end{figure}

Figure\,\ref{f_TAMS} shows a zoomed-in version of the Hertzsprung-Russell diagram for
both single and binary B supergiants from McEvoy et al. (2015). 
The position of the dark line indicates the position of the TAMS, with its location is determined by the value 
of the core overshooting parameter ($\alpha_{\rm ov}$) 
which is basically a ``free'' parameter (e.g. Vink et al. 2010; Brott et al. 2011) 
until astro-seismology on a large number of 
OB supergiants becomes available. The Brott et al. models employ a value of 
$\alpha_{\rm ov} = 0.335$, whilst the Geneva models (Georgy these proceedings) employ 
a smaller value. The VFTS results shown in Fig.\,\ref{f_TAMS} appear to suggest 
a {\it larger} value of $\alpha_{\rm ov}$ than 0.335. 

\begin{figure}
\begin{center}
\includegraphics[width=\textwidth]{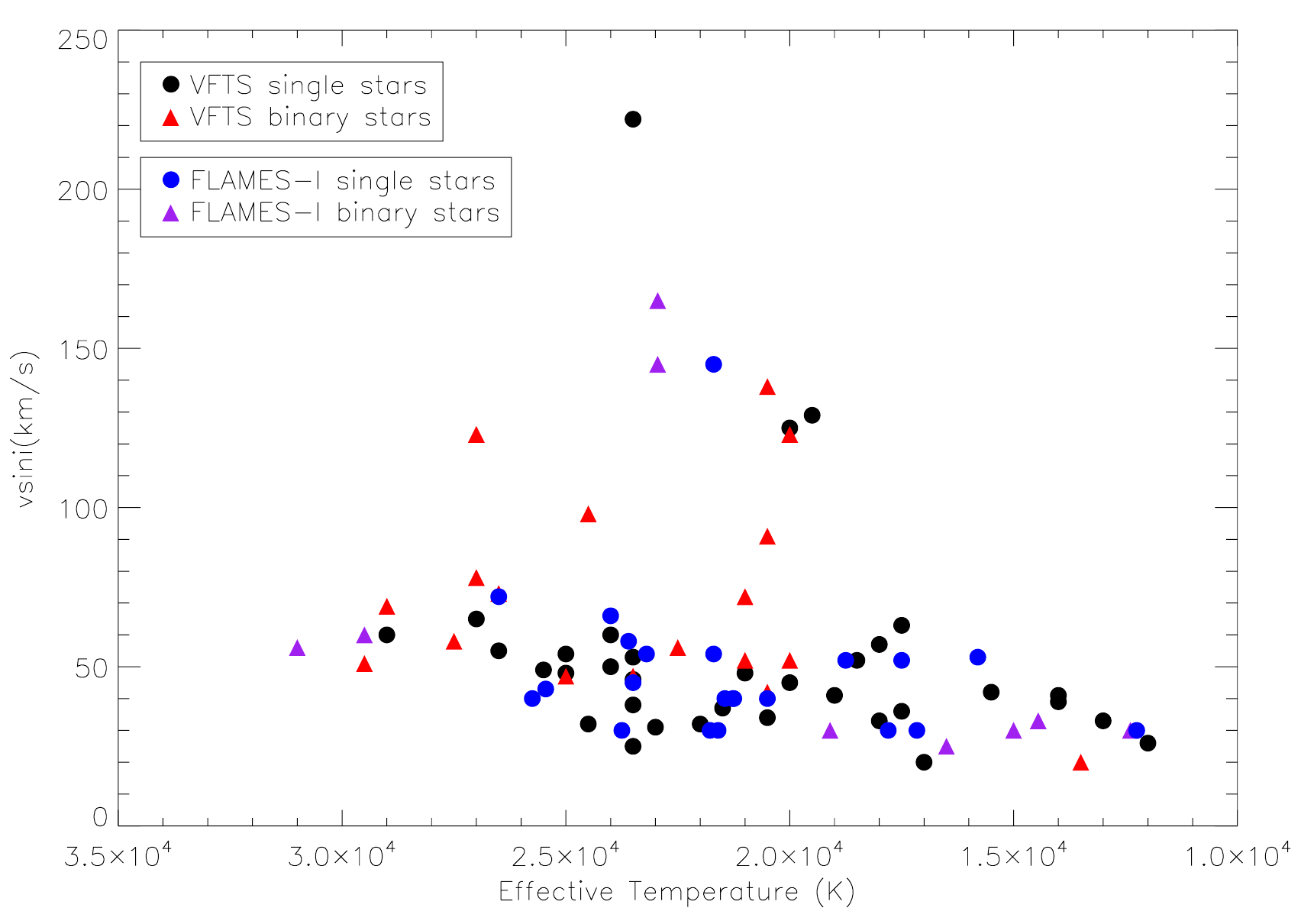}
\caption{Effective temperatures plotted against $v$ $\sin$ $i$ for FLAMES--I and VFTS B-type 
supergiants in the LMC (see Vink et al. 2010; McEvoy et al. 2015). The perceived lack of rapid rotators on the hot side 
of the diagram is not real, there are many rapidly rotating O-type stars; they are just not included here.}
\label{f_BSB}
\end{center}
\end{figure}

Larger $\alpha_{\rm ov}$ makes bi-stability braking (BSB; Vink et al. 2010; 
Keszthelyi et al. 2017) feasible, which we test by showing $v$ $\sin$ $i$ of both 
VFTS and previous FLAMES-I results (Hunter et al. 2008) versus 
$T_{\rm eff}$ in Fig.\,\ref{f_BSB}. 
Note the presence of another ``Region-of-Avoidance''\footnote{The perceived lack of rapid rotators on the hot side 
of the diagram is not real, there are many rapidly rotating O-type stars. These O-stars are just not included here.}, where 
rapidly-rotating ``cool'' (cooler than the bi-stability location of 20\,000\,K; Petrov et al. 2016) 
B supergiants are simply not observed. The reason for this avoidance below 20\,000\,K could either involve BSB, or it might be that the 
slowly rotating cool B supergiants are He-burning objects 
(due to post red-supergiant evolution or binarity). 

\section{The mass-loss rates}

\begin{figure}
\begin{center}
\includegraphics[width=\textwidth]{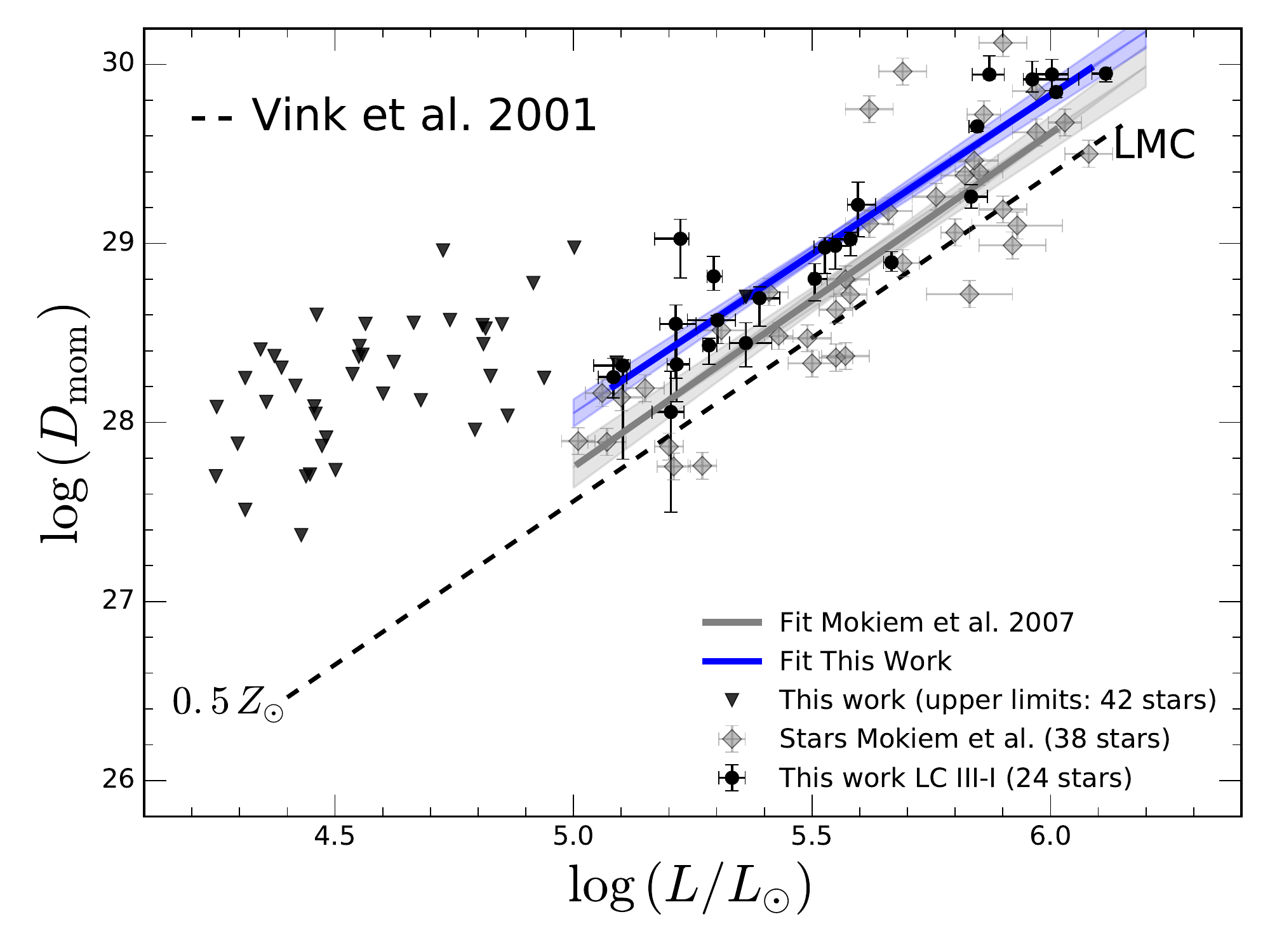}
\caption{Modified wind momentum ($D_{\rm mom}$) versus 
luminosity diagram from Ram\'{i}rez-Agudelo et al. (2017). 
The dashed lines indicate the theoretical predictions of Vink et al. (2001) 
for homogeneous winds. 
The empirical fit and Mokiem et al. (2007) (both for $L/L_{\odot} > 5.0$) are shown 
in shaded blue and gray bars, respectively. For stars with $L/L_{\odot} \leq\ 5.0$, only
upper limits could be constrained.}
\label{f_WLR}
\end{center}
\end{figure}

The mass-loss rates for O-type dwarfs were discussed by Sab{\'{\i}}n-Sanjuli{\'a}n et al. (2014; 2017), whilst 
those for the O giants and supergiants are plotted in the form of the 
wind-momentum-luminosity relation (WLR; Kudritzki \& Puls 2000; Puls et al. 2008) in Fig.\,\ref{f_WLR}. 
Interestingly, the empirical WLR lies above the theoretical WLR (of Vink et al. 2001). 
Usually a discrepancy between theoretical and empirical values would be interpreted 
such that the theoretical rates would be too low, but here it is different, as it is widely accepted that 
empirical modeling is more dependent on wind clumping and porosity than theory (see 
Muijres et al. 2011 for theoretical expectations).

Indeed, it is more likely that the empirical WLR is too high, as a result of wind clumping, 
which has not been included in the analysis. This would imply that the empirical WLR
would need to be lowered by a factor $\sqrt{D}$, where $D$ is the clumping factor, which 
is as yet uncertain. 
However, given the model-independent (from clumping \& porosity) transition mass-loss rate
(Vink \& Gr\"afener 2012; next Sect.) a value of $D \simeq 10$ (with a mass-loss rate and WLR reduction 
of $\sim$3) would bring the empirical WLR and theory in reasonable agreement. None of this means 
that the theoretical rates for lower mass-and-luminosity O stars need necessarily to be correct. Therefore, spectral analysis 
of large data-sets of O-stars including clumping \& porosity (Surlan et al. 2013; Sundqvist et al. 2014) 
are needed to provide definitive answers.

\begin{figure}
\begin{center}
\includegraphics[width=\textwidth]{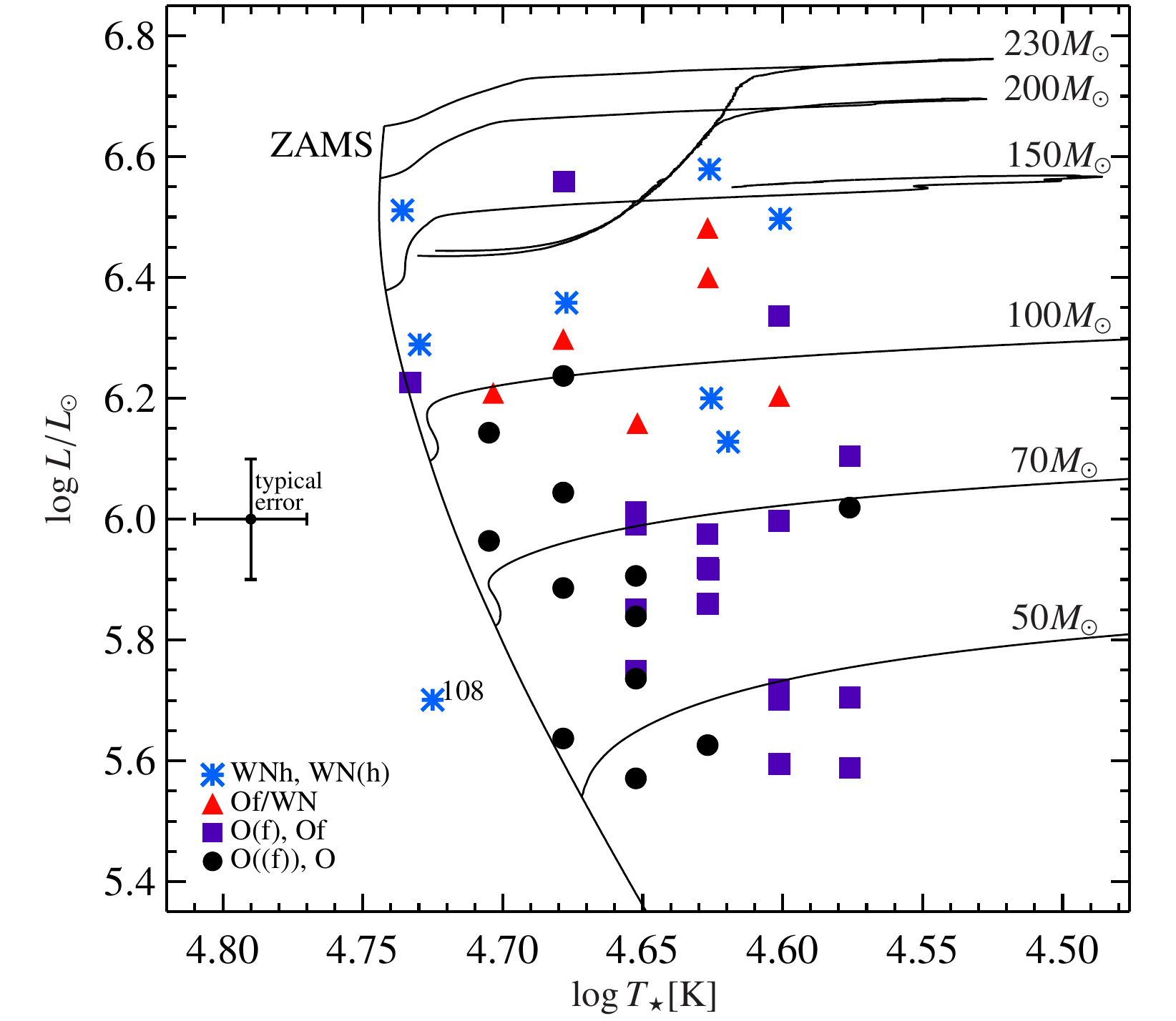}
\caption{Distribution of spectral types of the sample of Bestenlehner et al. (2014) 
in the HR-diagram. The
  different symbols indicate different stellar sub-classes. 
  Black lines indicate evolutionary tracks
  from K\"ohler et al. (2015) for an initial rotation rate of
  300\,km\,s$^{-1}$ and the location of the Zero-Age Main Sequence
  (ZAMS).}
\label{f_hrd_bonn}
\end{center}
\end{figure}

\section{Very Massive Stars}

\begin{figure}
\begin{center}
\includegraphics[width=\textwidth]{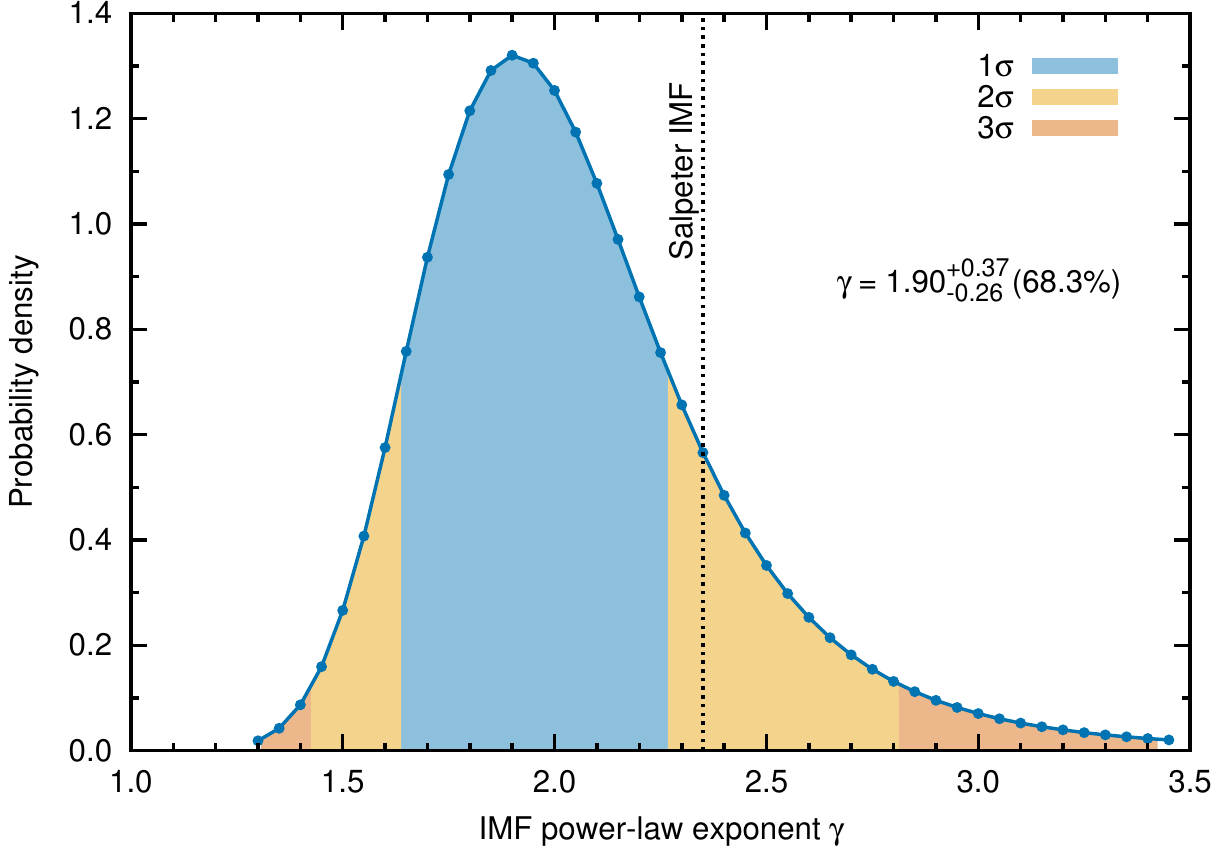}
\caption{Probability density distribution of the slope of the IMF (Schneider et al. 2017).}
\label{f_imf}
\end{center}
\end{figure}

The most massive stars in VFTS were analysed by Bestenlehner et al. (2014), 
plotted in the HRD of Fig.\,\ref{f_hrd_bonn}. Over-plotted 
are VMS evolutionary tracks and the location of the ZAMS.
The HRD shows the presence of 12 VMS (with $M$ $>$ 100 $M_{\odot}$; 
Vink et al. 2015), which enables us to derive the upper-IMF of 30 Dor for the first
time. 
Figure \ref{f_imf} compares the preferred value for the mass function to
that of Salpeter. It is found that the slope is different to that of Salpeter
(at $\sim$85\% confidence), and also that a Salpeter IMF cannot reproduce the larger 
number of massive stars above 30\,$M_{\odot}$ at $>$99\%
confidence (Schneider et al. 2017). 
As this result is obtained using the largest spectroscopic
data-set ever obtained, and analysed with the most sophisticated analysis tools, we consider this 
the most robust test to date. A top-heavy IMF would have major 
implications for the interpretation of spectral modelling of high-redshift galaxies, as well as the 
ionizing radiation and kinetic wind energy input into galaxies. Answers will strongly depend 
on the mass-loss rates of these VMS, as discussed next.

\begin{figure}
\begin{center}
\includegraphics[width=\textwidth]{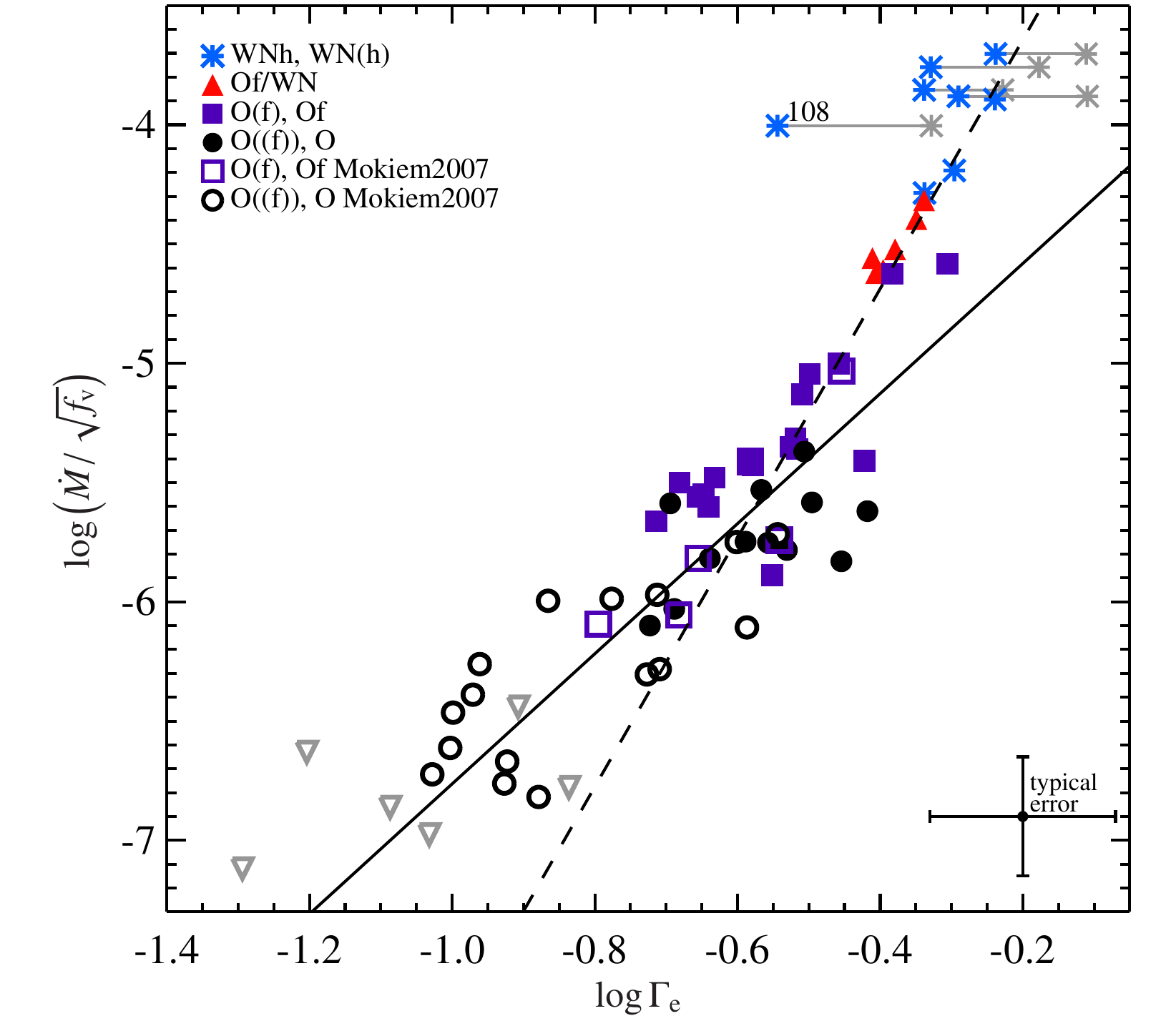}
\caption{Unclumped $\log \dot{M}$ vs.~$\log \Gamma_{\rm e}$. 
    Solid line: $\dot{M}-\Gamma_{\rm}$ relation for O stars. The
  different symbols indicate stellar sub-classes from Bestenlehner et al. (2014). 
  Dashed line: the steeper slope
  of the Of/WN and WNh stars. The {\it kink} occurs at $\log
  \Gamma_{\rm e} = -0.58$. The grey asterisks indicate the
    position of the stars with $Y > 0.75$ under the assumption of
    core He-burning. The grey upside down triangles are stars from
  Mokiem et al. (2007) which only have upper limits and
  are excluded from the fit. Note the presence of a kink, as predicted by Vink et al. (2011).}
\label{f_mdot_gamma}
\end{center}
\end{figure}

Figure\,\ref{f_mdot_gamma} shows VFTS results of the mass-loss rates 
of the most massive stars in 30 Dor (Bestenlehner et al. 2014). Whilst at relatively
low values of the Eddington value $\Gamma$, the slope of the empirical data is consistent
with that for O stars, those above the crossover point are not. 
Here the mass-loss rate kinks upwards, with a steeper slope. The winds have become 
optically thick, and show WR-like spectra. Also, above this critical $\Gamma$ 
point, the wind efficiency crosses unity, enabling 
a calibration of the absolute mass-loss rates for the first time (Vink \& Gr\"afener 2012). 
Moreover, Bestenlehner et al. (2014) found profound changes in the surface He abundances 
exactly coinciding with 
the luminosity threshold where mass loss is enhanced. This suggests that 
Of/WN and WNh stars are objects whose H-rich layers have been stripped
by enhanced mass-loss during their main-sequence life. 
Note that this mass-loss enhancement for VMS has not been included 
in most stellar evolution calculations, and this implies there will be many 
exciting surprises for extra-galactic applications of massive stars in the near future!

\section{Final Words}

The VFTS has conclusively shown that binaries are common in 30 Dor. With a corrected 
close-binary fraction of $\sim$50\% (Sana et al. 2013), we do not yet know whether this hints at 
a lower binary frequency at low metallicity, or it it is still consistent with the  
larger Galactic frequency of $\sim$70\% when evolutionary considerations are taken into account. 
Either way, we now know we require both single \& binary evolutionary models
to make progress. Another interesting finding is that there is a high-velocity tail present 
in single O-type supergiants (Ram\'{i}rez-Agudelo et al. 2013), which is not present in the 
spectroscopic binaries (Ram\'{i}rez-Agudelo et al. 2015). This suggests that binary interactions 
need to be accounted for to understand the underlying rotational distribution. 

The VFTS results also indicate that the main-sequence needs widening. This hints at a larger value for the core overshooting parameter than usually adopted. 
Finally, VMS up to at least 200$M_{\odot}$ are 
common in 30 Dor, but VMS mass-loss rates have been {\it under}estimated.

\end{document}